\documentclass[12pt,a4paper]{article}

\usepackage{times}
\usepackage{graphicx}
\usepackage{amssymb}
\usepackage{amsmath}

\newcommand{\abstracttitle}[1]{\parbox{\textwidth}
                              {\setlength{\baselineskip}{17pt}%
                               \begin{center}{\large\bfseries #1}\end{center}}}
\newcommand{\authors}[1]{\begin{center}#1\end{center}}
\newcommand{\addresses}[1]{\begin{center}#1\end{center}}

\newcommand{\cL}{{\cal L}}
\newcommand{\Tr}{{\rm Tr}}

\begin{document}

\abstracttitle{
PERTURBATIVE ANALYSIS OF NONEQUILIBRIUM STEADY STATES IN QUANTUM
SYSTEMS
}

\authors{{\bf Ayumu Sugita}}

\addresses{Department of Applied Physics, Osaka City University,\\
3-3-138 Sugimoto, Sumiyoshi-ku, Osaka, 558-8585, Japan}

\setlength{\parskip}{\bigskipamount}

\begin{abstract}

We study the nonequilibrium steady state (NESS) in a quantum system 
in contact with two heat baths at different temperatures.
We use a time-independent perturbative expansion with respect to
the coupling with the two heat baths to obtain the density matrix 
for the NESS. In particular, we show an explicit representation
of the density matrix for the reflection symmetric and weakly
nonequilibrium case. We also calculate the expectation value of
the energy current and show that the Kubo formula holds in this case. 

\end{abstract}

\section{Introduction}
Construction of nonequilibrium statistical mechanics is a challenging 
open problem in physics. Since nonequilibrium phenomena are so diverse,
probably it is impossible to make a theory which explains all 
nonequilibrium phenomena.
Then a natural first step would be statistical mechanics for nonequilibrium 
steady states (NESSs). The most ambitious goal in this direction is
to find a simple theoretical expression for the density matrix of
the NESSs. 

Although there are many theoretical frameworks to treat NESSs, it is
quite difficult to write down the density matrix explicitly.
For example, in the linear response theory [1] the density matrix for the
NESS is obtained in the long-time limit of 
the dynamics under an external field
\begin{eqnarray}
\hat{\rho}_{\rm NESS} &=& \hat{\rho}_{\rm eq} + 
\lim_{t\rightarrow\infty} \int_0^t dt' e^{-i(t-t')\hat{H}/\hbar}
\frac{1}{i\hbar}[\hat{H}_{\rm ext}, \hat{\rho}_{\rm eq}]e^{i(t-t')\hat{H}/\hbar},
\end{eqnarray} 
and the system has to be infinitely large. (Otherwise 
we obtain another equilibrium state.) 
Although this equation is
useful to calculate some nonequilibrium properties like
transport coefficients, $\hat{\rho}_{\rm NESS}$ itself
is very hard to calculate in this formalism. Most of formalisms
to treat NESS contain this kind of long time evolution, 
which makes it difficult
to calculate $\hat{\rho}_{\rm NESS}$.

In this paper, we consider a system with two heat baths.
We consider the stationary solution of a quantum master equation, and
calculate it explicitly using a perturbative expansion with respect to
the coupling parameter between the system and the heat baths.
In particular, in the reflection symmetric case we show an explicit
form of density matrix for the NESS in the weakly nonequilibrium regime.

\section{Equation of Motion}

\begin{figure}[h]
\begin{center}
\includegraphics*[height=5.0cm]{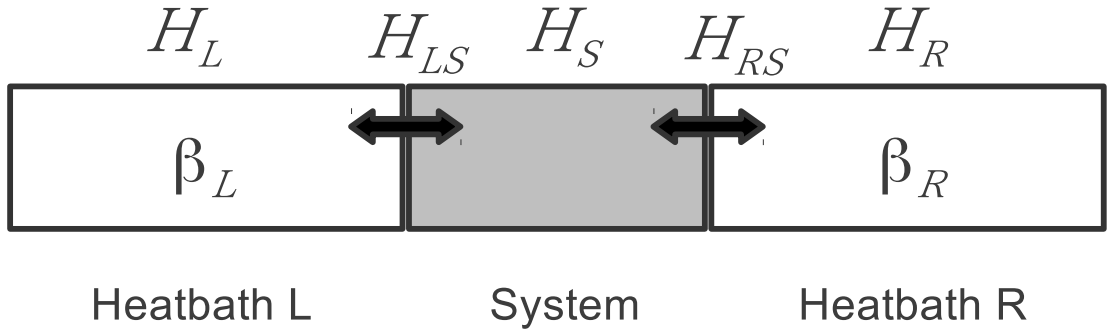}
\end{center}
\caption{The system is in contact with two heat baths at different inverse
temperatures $\beta_L$ and $\beta_R$.}
\label{setup}
\end{figure}

We start with the equation of motion for the total system:
\begin{eqnarray}
\frac{d}{dt}\hat{\rho}_{\rm tot}(t) &=& 
\frac{1}{i\hbar}[\hat{H}_{\rm tot},\hat{\rho}_{\rm tot}(t)],
\end{eqnarray}
where
\begin{eqnarray}
\hat{H}_{\rm tot} &=& \hat{H}_S + \hat{H}_B + u \hat{H}_{BS}.
\end{eqnarray}
Here $\hat{H}_S$, $\hat{H}_B$ and $\hat{H}_{BS}$ are 
the Hamiltonians of the system, 
the heat baths and the interactions, respectively.
We use $u$ as the perturbation parameter. 
In this paper, we consider a system with two heat baths:
\begin{eqnarray}
\hat{H}_B &=& \hat{H}_L + \hat{H}_R,\\
\hat{H}_{BS} &=& \hat{H}_{LS} + \hat{H}_{RS}. 
\end{eqnarray}
Here, $\hat{H}_L$ and $\hat{H}_R$ are the Hamiltonians for the left and
right reservoirs, respectively. We assume that the heat bath $\alpha$ ($\alpha
=L, R$) is in equilibrium with the inverse temperature $\beta_\alpha$
(See Fig. \ref{setup}).
The interaction Hamiltonians $\hat{H}_{LS}$ and $\hat{H}_{RS}$
can be written in the form
\begin{eqnarray}
\hat{H}_{LS} = \sum_j \hat{X}^L_j \hat{Y}^L_j,\;\;\;\;\;
\hat{H}_{RS} = \sum_j \hat{X}^R_j \hat{Y}^R_j
\end{eqnarray}
where $\hat{X}^{\alpha}_j$ acts on the system, 
and $\hat{Y}^L_j$ 
($Y^R_j$) acts on the left (right) heat bath. 
In the following we assume $\hat{X}^\alpha_j$
and $\hat{Y}^\alpha_j$ are Hermitian for simplicity. However, our main results
in this paper hold without this assumption.


We expand the density matrix up to $O(u^2)$,
trace out the heat bath variables, and apply the Markov
approximation. Then
we obtain the equation of motion for the system [2]
\begin{eqnarray}
\frac{d}{d t}\hat{\rho}(t) 
&=&
\frac{1}{i\hbar}[\hat{H}'_S, \hat{\rho}(t)] + 
u^2 \sum_{\alpha=L,R}\Gamma^\alpha \hat{\rho}(t).
\label{eq_motion}
\end{eqnarray}
Here, 
\begin{eqnarray}
\hat{H}'_S &\equiv& \hat{H}_S + 
u 
\sum_{\alpha=L,R}\sum_j \hat{X}^\alpha_j\langle \hat{Y}^\alpha_j\rangle
\end{eqnarray}
is the system Hamiltonian with the averaged interaction terms,
where $\langle \hat{A}^\alpha \rangle$ represents the average of $\hat{A}^\alpha$
with respect to the heat bath $\alpha$. Hereafter $\hat{H}'_S$ is denoted
as $\hat{H}_S$ for simplicity. $\Gamma^\alpha$ is the 
heat bath superoperator, whose explicit form is
\begin{eqnarray}
\Gamma^\alpha \hat{\rho}(t) &=& -\frac{1}{\hbar^2}\sum_{j,l}
\int_0^\infty dt' 
\left\{
\hat{X}^\alpha_j\hat{X}^\alpha_l(-t') \hat{\rho}(t)\Phi^\alpha_{jl}(t')
- 
\hat{X}^\alpha_j\hat{\rho}(t) \hat{X}^\alpha_l(-t')\Phi^\alpha_{lj}(-t')
\right. \nonumber \\
&&
\left.
+ \hat{\rho}(t) \hat{X}^\alpha_l(-t') \hat{X}^\alpha_j\Phi^\alpha_{lj}(-t')
- \hat{X}^\alpha_l(-t')\hat{\rho}(t) \hat{X}^\alpha_j \Phi^\alpha_{jl}(t')
\right\}
\label{Gamma}
\end{eqnarray}
Here, $\hat{X}(t) \equiv e^{-i\hat{H}_S t/\hbar}\hat{X}e^{i\hat{H}_S t/\hbar}$
represents the operator in the interaction picture. 
\begin{eqnarray}
\Phi^\alpha_{jl}(t) &\equiv& 
\langle \Delta \hat{Y}^\alpha_j(t) 
\Delta \hat{Y}_l^\alpha\rangle 
\end{eqnarray}
is a correlation function in the heat bath $\alpha$, where
\begin{eqnarray}
\Delta \hat{Y}^\alpha_j &
\equiv & \hat{Y}_j^\alpha - \langle \hat{Y}^\alpha_j\rangle.
\end{eqnarray}
Note that we used 
\begin{eqnarray}
\langle \Delta \hat{Y}^\alpha_j(t) 
\Delta \hat{Y}_l^\beta\rangle 
&=&
\delta_{\alpha\beta}
\langle \Delta \hat{Y}^\alpha_j(t) 
\Delta \hat{Y}_l^\alpha\rangle 
\end{eqnarray}
to derive Eq. (\ref{eq_motion}). Since we assumed $\hat{Y}^\alpha_j$'s are
Hermitian,
\begin{eqnarray}
\Phi_{jl}(t)^* &=& \Phi_{lj}(-t).
\end{eqnarray}

The heat bath superoperator (\ref{Gamma}) can be rewritten as
\begin{eqnarray}
\Gamma^\alpha &=& \Gamma_1^\alpha + \Gamma_2^\alpha, 
\end{eqnarray}
where
\begin{eqnarray}
\Gamma_1^\alpha \hat{\rho} &=&
-\frac{1}{2\hbar^2}\sum_{j,l}
\left(
[\hat{X}^\alpha_j, \hat{R}^\alpha_{jl}\hat{\rho}]
+
[\hat{X}^\alpha_j, \hat{R}^\alpha_{jl}\hat{\rho}]^\dagger
\right), \\
\Gamma_2^\alpha \hat{\rho} &=&
-\frac{i}{2\hbar^2}\sum_{j,l}
\left(
[\hat{X}^\alpha_j, \hat{W}^\alpha_{jl}\hat{\rho}]
-
[\hat{X}^\alpha_j, \hat{W}^\alpha_{jl}\hat{\rho}]^\dagger
\right). 
\end{eqnarray}
The operators $\hat{R}^\alpha_{jl}$ and $\hat{W}^\alpha_{jl}$ are defined as
\begin{eqnarray}
\langle E_p| \hat{R}^\alpha_{jl}|E_q\rangle &=& \langle E_p|\hat{X}^\alpha_l|E_q\rangle
\tilde{\Phi}_{jl}(\omega_{pq}),\\
\langle E_p|\hat{W}^\alpha_{jl}|E_q\rangle &=& \langle E_p|\hat{X}^\alpha_l|E_q\rangle
\tilde{\Psi}_{jl}(\omega_{pq}),
\end{eqnarray}
where $|E_i\rangle$ is an energy eigenvector of the system with
eigenenergy $E_i$, 
$\omega_{pq}\equiv (E_p - E_q)/\hbar$ and
\begin{eqnarray}
\tilde{\Phi}^\alpha_{jl}(\omega) &=& \int_{-\infty}^\infty dt e^{-i\omega t}\Phi^\alpha_{jl}(t)\\
\tilde{\Psi}^\alpha_{jl}(\omega) &=& {\cal P}\int_{-\infty}^\infty
\frac{d\omega'}{\pi}\frac{\Phi^\alpha_{jl}(\omega')}{\omega'-\omega}.
\end{eqnarray}
Here ${\cal P}$ denotes the Cauchy principal value. Note that
\begin{eqnarray}
\tilde{\Phi}^\alpha_{jl}(\omega)^*
&=& 
\tilde{\Phi}^\alpha_{lj}(\omega),\\
\tilde{\Psi}^\alpha_{jl}(\omega)^*
&=& 
\tilde{\Psi}^\alpha_{lj}(\omega).
\end{eqnarray}

The correlation functions satisfy the Kubo-Martin-Schwinger (KMS) condition
\begin{eqnarray}
\tilde{\Phi}^\alpha_{jl}(-\omega) &=& e^{\beta_\alpha \hbar\omega}
\tilde{\Phi}^\alpha_{lj}(\omega),
\end{eqnarray}
which is equivalent to the following operator identity:
\begin{eqnarray}
\hat{R}^{\alpha \dagger}_{jl} &=& 
e^{\beta_\alpha \hat{H}_S} \hat{R}^\alpha_{lj} e^{-\beta_\alpha \hat{H}_S}. 
\end{eqnarray}
Using this identity it is easy to show that
\begin{eqnarray*}
\Gamma^\alpha_1 e^{-\beta_\alpha \hat{H}_S} = 0,
\end{eqnarray*}
which guarantees the existence of the equilibrium solution
for Eq. (\ref{eq_motion}) when $\beta_L = \beta_R$.

\section{Perturbative expansion}
We put $\rho = {\rm const.}$ in the equation of motion (\ref{eq_motion}).
Then we have the equation for the steady state
\begin{eqnarray}
\cL_0 \hat{\rho} + v \cL_1 \hat{\rho} = 0,
\label{eq_ness}
\end{eqnarray}
where
\begin{eqnarray}
\cL_0 \hat{\rho} &=& \frac{1}{i\hbar}[\hat{H}_S, \hat{\rho}],\\
\cL_1 \hat{\rho} &=& \Gamma^L \hat{\rho} + \Gamma^R\hat{\rho},
\end{eqnarray}
and $v \equiv u^2$. We expand $\hat{\rho}$ with respect to
$v$:
\begin{eqnarray}
\hat{\rho} &=& \hat{\rho}_0 + v \hat{\rho}_1 + v^2 \hat{\rho}_2 + \dots
\end{eqnarray}
Then we obtain a series of equations
\begin{eqnarray}
\cL_0 \hat{\rho}_0 &=& 0,\label{0th}\\
\cL_0 \hat{\rho}_1 + \cL_1 \hat{\rho}_0 &=& 0,\label{1st}\\
\cL_0 \hat{\rho}_2 + \cL_1 \hat{\rho}_1 &=& 0, \label{2nd}\\
\vdots && 
\end{eqnarray}

\subsection{Separation of diagonal and off-diagonal parts}

In a normal perturbation theory, we can determine
$\hat{\rho}_i$ step by step starting from the 0th
order solution $\hat{\rho}_0$.
In this case, however, the 0th order equation (\ref{0th})
is degenerate, and any diagonal density matrices in the energy
representation satisfy it. Therefore we cannot fix
the 0th order term $\hat{\rho}_0$ from the 0th order equation
(\ref{0th}). 

To handle this problem, we introduce a projection 
superoperator $P$, which is defined by
\begin{eqnarray}
P |E_i\rangle \langle E_j| &=& 
\left\{
\begin{array}{cc}
|E_i\rangle \langle E_i| & (i=j) \\
0 & (i \ne j)
\end{array}
\right. .
\end{eqnarray}
Namely, $P$ is the projection to the diagonal part.
We also define $Q\equiv 1-P$, which is the projection
to the off-diagonal part.

Hereafter we assume that $\hat{H}_S$ is non-degenerate.
Then the 0th order equation (\ref{0th}) means that
$\hat{\rho}_0$ is diagonal.
Since $\cL_0$ satisfies 
\begin{eqnarray}
P \cL_0 = \cL_0 P = 0,
\end{eqnarray}
we obtain
\begin{eqnarray}
P\cL_1 \hat{\rho}_0 = P\cL_1 P \hat{\rho}_0 = 0
\label{0th_diag}
\end{eqnarray}
from the 1st order equation (\ref{1st}).
Eq. (\ref{0th_diag}) means $P\cL_1 P$ has a zero eigenvalue,
and we assume that it is non-degenerate.
Then Eq. (\ref{0th_diag}) 
determines the 0th order term $\hat{\rho}_0$ uniquely.

The unperturbed Liouvillian $\cL_0$ acts on the density matrix as
\begin{eqnarray}
\left(\cL_0 \hat{\rho}\right)_{jk} &=& 
\frac{1}{i\hbar}\left([\hat{H}_S, \hat{\rho}]\right)_{jk}\\
&=& 
\frac{E_j - E_k}{i\hbar}\rho_{jk},
\end{eqnarray}
where $A_{jk}\equiv \langle E_i|\hat{A}|E_k\rangle$ denotes
a matrix element in the energy representation. 
$\cL_0$ does not have its inverse because it has zero
eigenvalues. Nevertheless we can define  
its inverse in the off-diagonal subspace: 
\begin{eqnarray}
\left((Q\cL_0Q)^{-1}\hat{\rho}\right)_{jk}
&=& 
\frac{i\hbar}{E_j-E_k}\rho_{jk}.
\end{eqnarray}
Then from (\ref{1st}) we obtain the off-diagonal part of the first order term
\begin{eqnarray}
Q\hat{\rho}_1 &=& - (Q\cL_0 Q)^{-1} \cL_1 \hat{\rho_0}.
\end{eqnarray}

The diagonal part of the second order equation (\ref{2nd}) can be rewritten as
\begin{eqnarray}
P \cL_1 (P + Q) \hat{\rho}_1 &=& 0.
\label{1st_diag}
\end{eqnarray}
$P\cL_1 P$ has a zero eigenvalue, and the corresponding eigenvector is
$\hat{\rho}_0$. Therefore the general solution of (\ref{1st_diag}) is
\begin{eqnarray}
P\hat{\rho}_1 &=& - (P\cL_1P)^{'-1}P\cL_1 Q\hat{\rho}_1 + \gamma \hat{\rho}_0,
\end{eqnarray}
where $(P\cL_1P)^{'-1}$ is the inverse of $P\cL_1 P$ in the subspace
spanned by non-zero eigenvectors of $P\cL_1P$, and $\gamma$ is
a number determined by the normalization condition $\Tr \hat{\rho}_1 = 0$.

In the same procedure we can determine $Q\hat{\rho}_2$, $P\hat{\rho}_2$,
$Q\hat{\rho}_3$ and so on. These higher order terms, however, may be
physically irrelevant because Eq. (\ref{eq_ness}) was derived
from the approximation up to the first order of $v$.

\subsection{Perturbative expansion with respect to $\Delta\beta$}

The pertubative solution we have obtained in the previous subsection
is still very formal 
because we do not know the explicit form of the 0th order term $\rho_0$.
In this and following subsections we try to find a more explicit
solution by expanding $\hat{\rho}$ with respect to $\Delta\beta$,
the inverse temperature difference between the two heat bath.

We put 
\begin{eqnarray}
\beta_L &=& \beta - \frac{\Delta\beta}{2},\\
\beta_R &=& \beta + \frac{\Delta\beta}{2}.
\end{eqnarray}
Then we expand the heat bath superoperators and the density matrix as
\begin{eqnarray}
\Gamma^L(\beta_L) &=& 
\Gamma^L (\beta) - \frac{\Delta\beta}{2}\partial_\beta\Gamma^L(\beta)
+ O(\Delta\beta^2),\\   
\Gamma^R(\beta_R) &=& 
\Gamma^R (\beta) + \frac{\Delta\beta}{2}\partial_\beta\Gamma^L(\beta)
+ O(\Delta\beta^2),\\   
\end{eqnarray}
\begin{eqnarray}
\hat{\rho} &=&
\hat{\rho}_{00} + \Delta\beta \hat{\rho}_{01} + v
(\hat{\rho}_{10} + \Delta\beta \hat{\rho}_{11})
+ O(v^2) + O(\Delta\beta^2).
\end{eqnarray}
We obtain an equation for each order $O(v^n \Delta\beta^m)$:
\begin{eqnarray}
O(1): && \cL_0 \hat{\rho}_{00} = 0,\label{00}\\
O(\Delta\beta): && \cL_0 \hat{\rho}_{01} = 0,\label{01}\\
O(v): && \cL_0 \hat{\rho}_{10} + (\Gamma^L + \Gamma^R)
\hat{\rho}_{00}= 0,\label{10}\\
O(v \Delta\beta): && 
\cL_0 \hat{\rho}_{11} + (\Gamma^L + \Gamma^R)\hat{\rho}_{01} +
\frac{1}{2}(-\partial_\beta\Gamma^L + \partial_\beta \Gamma^R)\hat{\rho}_{00} = 0.
\label{11}   
\end{eqnarray}
Note that $\Gamma^\alpha$ and $\partial_\beta\Gamma^\alpha$
in the above equations are evaluated at the inverse temperature $\beta$.

Eqs. (\ref{00}) and (\ref{01}) mean that $\hat{\rho}_{00}$
and $\hat{\rho}_{01}$ are diagonal.
Since $\Gamma_2^\alpha$ satisfies
\begin{eqnarray}
P\Gamma^\alpha_2 P = 0,
\end{eqnarray}
we obtain
\begin{eqnarray}
P(\Gamma^L_1 + \Gamma^R_1) \hat{\rho}_{00} &=& 0
\end{eqnarray}
by applying $P$ to (\ref{10}). It has the equilibrium solution
\begin{eqnarray}
\hat{\rho}_{00} &=& \frac{1}{Z}e^{-\beta \hat{H}_S},
\end{eqnarray}
where $Z$ is the partition function.
Then from (\ref{10}) we obtain
\begin{eqnarray}
Q\hat{\rho}_{10} &=& - 
\frac{1}{Z}(Q\cL_0 Q)^{-1}(\Gamma^L_2 + \Gamma^R_2)e^{-\beta \hat{H}_S}.
\end{eqnarray}

\subsection{Symmetric case}
By applying $P$ to (\ref{11}) we obtain
\begin{eqnarray}
P(\Gamma^L_1 + \Gamma^R_1) \hat{\rho}_{01}
+
\frac{1}{2}P(-\partial_\beta\Gamma_1^L + \partial_\beta \Gamma_1^R)
\hat{\rho}_{00} = 0.
\label{11'}
\end{eqnarray}
In principle, $\hat{\rho}_{01}$ is determined by solving this equation.
However, the inverse of $P(\Gamma^L_1 + \Gamma^R_1)P$ is hard to
calculate analytically.

Here we assume that the system and the heat baths are reflection symmetric.
More precisely, we assume that $\hat{\Pi}\hat{H}_{\rm tot} \hat{\Pi} =
\hat{H}_{\rm tot}$, where $\hat{\Pi}$ is the parity operator
which satisfies $\hat{\Pi}^2 = 1$. Then we have
\begin{eqnarray}
\hat{\Pi} \hat{H}_S \hat{\Pi} &=& \hat{H}_S,\\
\hat{\Pi} \hat{X}^L_j \hat{\Pi} &=& \hat{X}^R_j,\\
\hat{\Pi} \hat{R}^L_{jl} \hat{\Pi} &=& \hat{R}^R_{jl},\label{R}\\
\hat{\Pi} \hat{W}^L_{jl} \hat{\Pi} &=& \hat{W}^R_{jl}.\label{W}
\end{eqnarray}
Note that 
the operators
are evaluated at the same inverse temperature $\beta$
in Eqs. (\ref{R}) and (\ref{W}). 

Then let us consider the second term of
Eq. (\ref{11'}). Since $\rho_{00}$ is symmetric, we have
\begin{eqnarray}
\hat{\Pi} \partial_\beta \Gamma_1^L \rho_{00} \hat{\Pi}
&=&
\partial_\beta \Gamma_1^R \rho_{00}\\ 
\hat{\Pi} \partial_\beta \Gamma_1^R \rho_{00} \hat{\Pi}
&=&
\partial_\beta \Gamma_1^L \rho_{00}. 
\end{eqnarray}
A diagonal element in the second term of (\ref{11'})
is
\begin{eqnarray}
\langle E_p | (- \partial_\beta \Gamma_1^L +\partial_\beta \Gamma_1^R) \hat{\rho}_{00}|E_p\rangle
&=&
\langle E_p |
\hat{\Pi} (- \partial_\beta \Gamma_1^L +\partial_\beta \Gamma_1^R) \hat{\rho}_{00}\hat{\Pi}
|E_p\rangle\\
&&
\left(\because \hat{\Pi}|E_p\rangle = \pm |E_p\rangle \right)\\
&=&
\langle E_p | (- \partial_\beta \Gamma_1^R +\partial_\beta \Gamma_1^L) \hat{\rho}_{00}|E_p\rangle\\
&=&
- \langle E_p | (- \partial_\beta \Gamma_1^L +\partial_\beta \Gamma_1^L) \hat{\rho}_{00}|E_p\rangle.
\end{eqnarray}
Hence
\begin{eqnarray}
\langle E_p | (- \partial_\beta \Gamma_1^L +\partial_\beta \Gamma_1^L) \hat{\rho}_{00}|E_p\rangle
&=& 0
\end{eqnarray}
and the second term of (\ref{11'}) vanishes. Then we have
\begin{eqnarray}
P (\Gamma^L_1 + \Gamma^R_1) \hat{\rho}_{01} = 0,
\end{eqnarray}
whose solution is
\begin{eqnarray}
\hat{\rho}_{01} &\propto & \hat{\rho}_{00} = \frac{1}{Z}e^{-\beta \hat{H}_S}.
\end{eqnarray}
To keep the normalization condition $\Tr \hat{\rho}=1$, we should put
\begin{eqnarray}
\hat{\rho}_{01} &=& 0.
\end{eqnarray}
Then we obtain the lowest order nonequilibrium term
\begin{eqnarray}
Q \hat{\rho}_{11} &=& - \frac{1}{2Z} (Q\cL_0 Q)^{-1}
Q(-\partial_\beta \Gamma^L + \partial_\beta \Gamma^R)e^{-\beta \hat{H}_S}.
\label{main}
\end{eqnarray}
from (\ref{11}). 
This is our main result.
Note that diagonal elements do not contribute to
nonequilibrium properties like the energy current and
the temperature gradient.

\section{Energy current}

\subsection{Energy current operator}
Let us consider the energy current going through the system.
We divide the system into two parts. Then the system Hamiltonian
is
\begin{eqnarray}
\hat{H}_S &=& \hat{H}_l + \hat{H}_i + \hat{H}_r,
\end{eqnarray}
where $\hat{H}_l$ and $\hat{H}_r$ are the Hamiltonians for
the left and right parts of the system, respectively, and
$\hat{H}_i$ is the interaction between them. 
Note that $[\hat{H}_l, \hat{H}_r] = 0$.
The energy current
which goes from the left part to the right part can be
defined as the energy loss of the left part:
\begin{eqnarray}
\hat{J}_l &\equiv& - \dot{\hat{H}}_l 
= - \frac{1}{i\hbar}[\hat{H}_l, \hat{H}_S]  
= - \frac{1}{i\hbar}[\hat{H}_l, \hat{H}_i].
\end{eqnarray}
It is also possible to define another current operator
by the energy gain of the right part:
\begin{eqnarray}
\hat{J}_r &\equiv & \dot{\hat{H}}_r
= \frac{1}{i\hbar}[\hat{H}_r, \hat{H}_S]  
= \frac{1}{i\hbar}[\hat{H}_r, \hat{H}_i].
\end{eqnarray}

We can also define the energy current at a boundary between
the system and a heat bath. The total energy of the system changes as
\begin{eqnarray}
\frac{d}{dt} \langle \hat{H}_S \rangle
&=&
\Tr\left( \hat{H}_S \frac{d}{dt}\hat{\rho}\right) \\
&=&
\Tr \left\{ \hat{H}_S \left([\hat{H}_S, \hat{\rho}]
+ v \Gamma^L\hat{\rho} + v \Gamma^R \hat{\rho}
\right)
\right\}\\
&=&
v \Tr \left( \hat{H}_S \Gamma^L \hat{\rho}\right) 
+ 
v \Tr \left( \hat{H}_S \Gamma^R \hat{\rho}\right).
\end{eqnarray}
The left (right) term can be interpreted as the energy current 
at the left (right) boundary. Therefore we define two current operators
$\hat{J}_L$ and $\hat{J}_R$ so that the following relations hold.
\begin{eqnarray}
\langle \hat{J}_L \rangle &=& v \Tr \left( \hat{H}_S \Gamma^L \hat{\rho}\right) ,
\label{J_L1} \\
\langle \hat{J}_R \rangle &=& - v \Tr \left( \hat{H}_S \Gamma^R \hat{\rho}\right).
\end{eqnarray}
Then
\begin{eqnarray}
&&
\Tr ( \hat{J}_L \hat{\rho}) \\
&=& 
-\frac{v}{2\hbar^2}\sum_{jl}
\Tr \left\{ \hat{H}_S 
\left(
[\hat{X}^L_j, \hat{R}^L_{jl}\hat{\rho}]
+
[\hat{X}^L_j, \hat{R}^L_{jl}\hat{\rho}]^\dagger
\right)
+ i
\left(
[\hat{X}^L_j, \hat{W}^L_{jl}\hat{\rho}]
-
[\hat{X}^L_j, \hat{W}^L_{jl}\hat{\rho}]^\dagger
\right)
\right\} \\
&=&
-\frac{v}{2\hbar^2}\sum_{jl}
\Tr \left\{  
\left(
[\hat{H}_S, \hat{X}^L_j] 
\left(\hat{R}^L_{jl} + i\hat{W}^L_{jl}\right)
+
\left(\hat{R}^L_{jl} + i\hat{W}^L_{jl}\right)^\dagger
[\hat{H}_S, \hat{X}^L_j]^\dagger 
\right)
\hat{\rho}
\right\}.
\end{eqnarray}
Hence
\begin{eqnarray}
\hat{J}_L 
&=&
-\frac{v}{2\hbar^2}\sum_{jl}
\left\{
[\hat{H}_S, \hat{X}^L_j] 
\left(\hat{R}^L_{jl} + i\hat{W}^L_{jl}\right)
+
\left(\hat{R}^L_{jl} + i\hat{W}^L_{jl}\right)^\dagger
[\hat{H}_S, \hat{X}^L_j]^\dagger 
\right\}.
\label{J_L2}
\end{eqnarray}
In the same way we obtain
\begin{eqnarray}
\hat{J}_R 
&=&
\frac{v}{2\hbar^2}\sum_{jl}
\left\{
[\hat{H}_S, \hat{X}^R_j] 
\left(\hat{R}^R_{jl} + i\hat{W}^R_{jl}\right)
+
\left(\hat{R}^R_{jl} + i\hat{W}^R_{jl}\right)^\dagger
[\hat{H}_S, \hat{X}^R_j]^\dagger 
\right\}.
\end{eqnarray}
In the steady state all current operators should have the same
expectation value.

\subsection{Expectation value}

Let us consider the expectation value of a energy current operator
in the system. In our perturbation theory, $Q\hat{\rho}_{11}$ is 
the leading nonequilibrium term. Therefore we evaluate 
\begin{eqnarray}
\Tr (\hat{J}_l Q\hat{\rho}_{11})
&=&
\frac{1}{2 i\hbar}\left\{[\hat{H}_l, \hat{H}_S]
(Q\cL_0 Q)^{-1}(-\partial_\beta\Gamma^L + \partial_\beta \Gamma^R)
\hat{\rho}_{00}
\right\}
\\
&=&
-\frac{1}{2}\left\{(\cL_0 \hat{H}_l)
(Q\cL_0 Q)^{-1}Q(-\partial_\beta\Gamma^L + \partial_\beta \Gamma^R)
\hat{\rho}_{00}
\right\}.
\end{eqnarray}
Note that
\begin{eqnarray}
\Tr \left\{(\cL_0 \hat{A})(Q\cL_0^{-1}Q)^{-1}Q\hat{B}\right\}
&=&
-\Tr (\hat{A}Q\hat{B})
\end{eqnarray}
holds in general, because
\begin{eqnarray}
\Tr \left\{(\cL_0 \hat{A})(Q\cL_s^{-1}Q)^{-1}Q\hat{B}\right\}
&=&
\sum_{l,m} \langle E_l|\cL_0 \hat{A} |E_m\rangle 
\langle E_m |(Q\cL_0^{-1}Q)^{-1}Q\hat{B}|E_l\rangle\\
&=&
\sum_{l\ne m} (E_l - E_m)\langle E_l|\hat{A}|E_m\rangle 
\frac{1}{E_m - E_l}\langle E_m |\hat{B}|E_l\rangle\\
&=&
- \sum_{l\ne m} \langle E_l|\hat{A}|E_m\rangle 
\langle E_m |\hat{B}|E_l\rangle\\
&=&
- \Tr (\hat{A}Q\hat{B}).
\end{eqnarray}
Hence
\begin{eqnarray}
\Tr (\hat{J}_l Q \hat{\rho}_{11})
&=&
\frac{1}{2}\Tr \left\{\hat{H}_l
Q(-\partial_\beta\Gamma^L + \partial_\beta \Gamma^R)
\hat{\rho}_{00}
\right\}\\
&=&
\frac{1}{2}\Tr \left\{\hat{H}_l
(-\partial_\beta\Gamma^L + \partial_\beta \Gamma^R)
\hat{\rho}_{00}
\right\}\\
&&
\left(\because P(-\partial_\beta\Gamma^L + \partial_\beta \Gamma^R)
\hat{\rho}_{00} = 0
\right).
\end{eqnarray}
Each term in $\partial_\beta\Gamma^\alpha\hat{\rho}_{00}$
has the form $[\hat{X}^\alpha_j, \partial_\beta \hat{Z}_{jl}\hat{\rho}_{00}]$
($Z = R, W$). Then
\begin{eqnarray}
\Tr \left(\hat{H}_l[\hat{X}^\alpha_j, \partial_\beta \hat{Z}_{jl}\hat{\rho}_{00}]\right)
&=&
\Tr \left([\hat{H}_l, \hat{X}^\alpha_j] \partial_\beta \hat{Z}_{jl}\hat{\rho}_{00}\right),
\end{eqnarray}
which vanishes if $\alpha = R$. Hence
\begin{eqnarray}
\Tr (\hat{J}_l Q \hat{\rho}_{11})
&=&
-\frac{1}{2}\Tr \left(
\hat{H}_l\partial_\beta\Gamma^L \hat{\rho}_{00}
\right)\\
&=&
-\frac{1}{2}\Tr \left(
\hat{H}_S\partial_\beta\Gamma^L \hat{\rho}_{00}
\right) \label{j_l_exp}
\\
&=&
-\frac{1}{2}\Tr \left\{
\hat{H}_S\left(\partial_\beta\Gamma^L_1 + \partial_\beta\Gamma^L_2\right)\hat{\rho}_{00}
\right\}.
\end{eqnarray}
In the second line we used $[\hat{H}_l, \hat{X}_j] = [\hat{H}_S, \hat{X}_j]$.
With some algebra, one can easily show 
\begin{eqnarray}
\Tr \left( \hat{H}_S \partial_\beta\Gamma^L_2\hat{\rho}_{00}\right) = 0.
\end{eqnarray}
Since
\begin{eqnarray}
\Gamma_1^L (\beta) e^{-\beta \hat{H}_S} = 0
\end{eqnarray}
for any $\beta$,
\begin{eqnarray}
\partial_\beta \left(\Gamma_1^L (\beta) e^{-\beta \hat{H}_S}\right)
= 
\partial_\beta\Gamma_1^L (\beta) e^{-\beta \hat{H}_S}
+
\Gamma_1^L (\beta) \partial_\beta e^{-\beta \hat{H}_S}
= 0.
\end{eqnarray}
Therefore
\begin{eqnarray}
\partial_\beta \Gamma_1^L \hat{\rho}_{00} 
&=&
-\Gamma_1^L \hat{H}_S \hat{\rho}_{00} 
\end{eqnarray}
and
\begin{eqnarray}
\Tr (\hat{J}_l Q \hat{\rho}_{11})
&=&
\frac{1}{2}
\Tr \left(\hat{H}_S \Gamma_1^L \hat{H}_S \rho_{00}\right)\\
&=&
-\frac{1}{2\hbar^2}{\rm Re}
\sum_{jl}
\Tr \left(\hat{H}_S [\hat{X}^L_j, \hat{R}^L_{jl}\hat{H}_S \rho_{00}]\right)\\
&=&
-\frac{1}{2\hbar^2}{\rm Re}
\sum_{jl}
\Tr \left([\hat{H}_S, \hat{X}^L_j] \hat{R}^L_{jl}\hat{H}_S \rho_{00}\right)\\
&=&
-\frac{1}{4\hbar^2 Z}
\sum_{jl}\sum_{pq}
\left\{
E_p(E_p-E_q)\langle E_p|\hat{X}^L_j|E_q\rangle \langle E_q|\hat{X}^L_l| E_p\rangle
\Phi_{jl}(\omega_{qp}) e^{-\beta E_p}
\right.\nonumber\\
&&
+
\left.
E_p(E_p-E_q)\langle E_q|\hat{X}^L_j|E_p\rangle \langle E_p|\hat{X}^L_l| E_q\rangle
\Phi_{lj}(\omega_{qp}) e^{-\beta E_p}
\right\}\\
&=&
-\frac{1}{4\hbar^2 Z}
\sum_{jl}\sum_{pq}
\left\{
E_p(E_p-E_q)\langle E_p|\hat{X}^L_j|E_q\rangle \langle E_q|\hat{X}^L_l| E_p\rangle
\Phi_{jl}(\omega_{qp}) e^{-\beta E_p}
\right.\nonumber\\
&&
+
\left.
E_q(E_q-E_p)\langle E_p|\hat{X}^L_j|E_q\rangle \langle E_q|\hat{X}^L_l| E_p\rangle
\Phi_{lj}(\omega_{pq}) e^{-\beta E_q}
\right\}\\
&=&
-\frac{1}{4\hbar^2 Z}
\sum_{jl}\sum_{pq}
\left\{
(E_p-E_q)^2\langle E_p|\hat{X}^L_j|E_q\rangle \langle E_q|\hat{X}^L_l| E_p\rangle
\Phi_{jl}(\omega_{qp}) e^{-\beta E_p}
\right\}\\
&=&
-\frac{1}{4\hbar^2}\sum_{jl}\left\langle [\hat{H}_S, \hat{X}_j^L] 
[\hat{R}^L_{jl}, \hat{H}_S] \right\rangle_\beta.
\end{eqnarray}
In the last line, the expectation value is evaluated for
the system in equilibrium at the inverse temperature $\beta$.

In the same way we obtain
\begin{eqnarray}
{\rm Tr}(\hat{J}_r Q\rho_{11}) &=&
-\frac{1}{4\hbar^2}\sum_{jl}\left\langle [\hat{H}_S, \hat{X}_j^R] 
[\hat{R}^R_{jl}, \hat{H}_S] \right\rangle_\beta.
\end{eqnarray}
Since we have assumed the reflection symmetry, we obtain
\begin{eqnarray}
\langle \hat{J}_l\rangle 
=
\langle \hat{J}_r\rangle
&=&
-\frac{1}{4\hbar^2}\sum_{jl}\left\langle [\hat{H}_S, \hat{X}_j^\alpha] 
[\hat{R}^\alpha_{jl}, \hat{H}_S] \right\rangle_\beta v \Delta\beta
+ O(v^2) + O(\Delta\beta^2).
\end{eqnarray}

We can also calculate the expectation values of the current operators
at the boundaries. Let us consider $\hat{J}_L$.
Since $\hat{J}_L$ is  $O(v)$, the
lowest order current comes from $O(v^0)$ terms of the density matrix.
Because $\hat{\rho}_{01} = 0$ in the symmetric case, we have
\begin{eqnarray}
\langle \hat{J}_L \rangle 
&=&
\Tr (\hat{J}_L \hat{\rho}_{00}) 
+ O(v^2) + O(\Delta\beta^2).
\end{eqnarray}  
Then, from Eq. (\ref{J_L1})
\begin{eqnarray}
\Tr (\hat{J}_L \hat{\rho}_{00}) 
&=&
v \Tr\left(\hat{H}_S \Gamma^L \rho_{00}\right).
\end{eqnarray}
Note that $\Gamma^L$ is evaluated at $\beta_L = \beta - \Delta\beta/2$
here. Substituting
\begin{eqnarray}
\Gamma_L(\beta_L) &=& \Gamma_L (\beta) - \frac{\Delta\beta}{2}\partial_\beta \Gamma_L(\beta)
\end{eqnarray}
we obtain
\begin{eqnarray}
\Tr (\hat{J}_L \hat{\rho}_{00}) 
&=&
- \frac{v \Delta\beta}{2} 
\Tr \left(\hat{H}_S \partial_\beta\Gamma^L \hat{\rho_{00}}\right),
\end{eqnarray}
which is equivalent to (\ref{j_l_exp}). Then we obtain the same current expectation
value again:
\begin{eqnarray}
\langle \hat{J}_L \rangle = \langle \hat{J}_R \rangle
&=&
-\frac{1}{4\hbar^2}
\sum_{jl}\left\langle [\hat{H}_S, \hat{X}_j^R] 
[\hat{R}^R_{jl}, \hat{H}_S] \right\rangle_\beta v \Delta\beta
+ O(v^2) + O(\Delta\beta^2).
\end{eqnarray}

\subsection{Kubo formula}
Let us consider the total system including the heat bath again.
The current operator for the energy coming from 
the heat bath L to the system is defined as
\begin{eqnarray}
\hat{J}'_{L} &=& \frac{d}{dt}(\hat{H}_S + u \hat{H}_{RS} + \hat{H}_R)\\
&=&
\frac{1}{i\hbar}[\hat{H}_S + u \hat{H}_{RS} + \hat{H}_R, \hat{H}_{\rm tot}]\\
&=&
\frac{1}{i\hbar}[\hat{H}_S, u \hat{H}_{LS}]\\
&=&
\frac{u }{i\hbar}\sum_j [\hat{H}_S, \hat{X}_j] \hat{Y}_j.
\end{eqnarray}
Then we define the correlation function
\begin{eqnarray}
C_L (t) &\equiv& \frac{1}{2}
\langle \hat{J}'_L(t) \hat{J}'_L + \hat{J}'_L \hat{J}'_L(t)\rangle.
\end{eqnarray}
The expectation value is evaluated for the equilibrium of the total system
with inverse temperature $\beta$. Since $\hat{J}'_L$ is $O(u^2)$,
we have
\begin{eqnarray}
C_L(t) &=& C^{(0)}(t) + O(u^3),
\end{eqnarray}
where
\begin{eqnarray}
C^{(0)}(t) = 
\frac{1}{2}\Tr \left\{\hat{\rho}_S^{\rm eq}\otimes \hat{\rho}_B^{\rm eq}
\left(
e^{i(\hat{H}_S+\hat{H}_B)t/\hbar} 
\hat{J}_L'
e^{-i(\hat{H}_S+\hat{H}_B)t/\hbar} 
\hat{J}_L'
+
\hat{J}_L'
e^{i(\hat{H}_S+\hat{H}_B)t/\hbar} 
\hat{J}_L'
e^{-i(\hat{H}_S+\hat{H}_B)t/\hbar} 
\right)
\right\}.
\end{eqnarray}
Here, $\hat{\rho}_S^{\rm eq}$ 
and $\hat{\rho}_B^{\rm eq}$ represents the equilibrium state
of the system and the heat baths, respectively.
Then we have
\begin{eqnarray}
C_L^{(0)}(t) &=& 
-\frac{u^2}{2\hbar^2}\sum_{jl}
\left\{
\left\langle [H_S, \hat{X}_j(t)][H_S, \hat{X}_l] \right\rangle_\beta
\Phi_{jl}(t) 
+ 
\left\langle [H_S, \hat{X}_j][H_S, \hat{X}_l(t)] \right\rangle_\beta
\Phi_{jl}(-t)
\right\}.
\end{eqnarray}
and
\begin{eqnarray}
\int_0^\infty dt\, C_L^{(0)}(t)
&=&
-\frac{u^2}{2Z\hbar^2}\sum_{jl}
\sum_{p,q}
e^{-\beta E_p}
(E_p - E_q)^2 \langle E_p|\hat{X}_j |E_q\rangle \langle E_q |\hat{X}_l |E_r\rangle
\nonumber\\
&&
\times
\int_0^\infty dt 
\left\{
e^{-i\omega_{qp}t}\Phi_{jl}(t) + e^{i\omega_{qp}t}\Phi_{jl}(-t)
\right\}\\
&=&
-\frac{u^2}{2Z\hbar^2}\sum_{jl}
\sum_{p,q}
e^{-\beta E_p}(E_p - E_q)^2 \langle E_p|\hat{X}_j |E_q\rangle 
\langle E_q |\hat{X}_l |E_r\rangle
\tilde{\Phi}_{jl}(\omega_{qp}) \\
&=&
-\frac{v}{2\hbar^2}
\sum_{jl}\left\langle [\hat{H}_S, \hat{X}_j^R] 
[\hat{R}^R_{jl}, \hat{H}_S] \right\rangle_\beta . 
\end{eqnarray}
Therefore, in the lowest order, the current is written as
\begin{eqnarray}
\langle \hat{J}_L \rangle &=& \frac{\Delta\beta}{2} 
\int_0^t dt\, C_L (t), 
\label{kubo}
\end{eqnarray}
which is the Kubo formula [3], or the fluctuation-dissipation theorem, 
in this case.
Note that $\Delta\beta/2$ is the inverse temperature difference at
the boundary. 

In spite of the formal similarity, 
physical content of Eq. (\ref{kubo}) is quite different from
the original Kubo formula [1]. For example, the transport coefficient
contains the information of the heat baths, though the original one does not.

\section{Summary}
We have calculated the density matrix for the NESS using the time-independent
perturbation theory. Our main result is Eq. (\ref{main}), which
is an explicit expression for the density matrix for the NESS in
the reflection symmetric setting. We have also calculated the
expectation value of the energy current and shown that
the Kubo formula holds in this case.

\section*{Acknowledgement}
The author would like to thank T. Yuge and T. Monnai for helpful
comments and discussions

\vspace*{4ex}
\setlength{\parskip}{\smallskipamount}

\noindent
\underline{References}
\frenchspacing
\begin{enumerate}
\setlength{\itemindent}{-0.7em}
\setlength{\itemsep}{0ex}
\item R. Kubo, J. Phys. Soc. Jpn. {\bf 12},570 (1957).

\item R. Kubo, M. Toda and N. Hashitsume,
{\it Statistical Physics II}, Springer.

\end{enumerate}

\end{document}